# Analytical Device-Physics Framework for Non-Planar Solar Cells

T. Kirkpatrick[*], M. J. Burns, M. J. Naughton

Department of Physics, Boston College, 140 Commonwealth Avenue, Chestnut Hill, Massachusetts 02467, USA

[*]Present Address: Department of Mechanical Engineering, Massachusetts Institute of Technology, 77 Massachusetts Avenue, Cambridge, Massachusetts 02139, USA

**ABSTRACT**

Non-planar solar-cell devices have been promoted as a means to enhance current collection in absorber materials with charge-transport limitations. This work presents an analytical framework for assessing the ultimate performance of non-planar solar-cells based on materials and geometry. Herein, the physics of the *p-n* junction is analyzed for low-injection conditions, when the junction can be considered spatially separable into quasi-neutral and space-charge regions. For the conventional planar solar cell architecture, previously established one-dimensional expressions governing charge carrier transport are recovered from the framework established herein. Space-charge region recombination statistics are compared for planar and non-planar geometries, showing variations in recombination current produced from the space-charge region. In addition, planar and non-planar solar cell performance are simulated, based on a semi-empirical expression for short-circuit current, detailing variations in charge carrier transport and efficiency as a function of geometry, thereby yielding insights into design criteria for solar cell architectures. For the conditions considered here, the expressions for generation rate and total current are shown to universally govern any solar cell geometry, while recombination within the space-charge region is shown to be directly dependent on the geometrical orientation of the *p-n* junction.

## I. INTRODUCTION

Contemporary solar cell designs are based on a planar geometry, for which charge carrier separation and photon absorption are approximately one-dimensional within the device. In addition to factors such as material and manufacturing costs [1-3], novelty of non-planar solar cell architectures is grounded in the idea that some non-planar devices decouple optical and electronic path lengths [4] and, therefore, offer opportunities to alter the competing roles of charge carrier collection and recombination within a device, which limit efficiency for planar cells with low charge carrier mobility and lifetime. In recent years, a number of unconventional, non-planar solar cell designs have been proposed, and some experimentally fabricated [4-13], in efforts to increase energy conversion efficiencies. To date, however, the planar solar cell architecture still holds all efficiency records over its non-planar counterparts [14]. While there is substantial information describing analytical device physics of planar solar cells [15-24], a comparatively small amount of literature is available describing analytical charge carrier transport properties for non-planar solar cell devices [25-30]. By detailing the device physics of a "geometrically generalized" solar cell, devices of various geometrical architectures are modeled congruently to ascertain under what conditions non-planar architectural configurations improve efficiency.

Our aim is to develop a simple framework for analytically calculating solar cell current as a function of voltage for a geometrically generalized *p-n* junction solar cell; analogous to the model employed to analytically calculate current as a function of voltage for a planar *p-n* junction solar cell. The purpose in doing so is to provide a better practical understanding of charge carrier transport for non-planar solar cells, and to explain how geometry, alone, can easily, and significantly, alter solar cell performance. Previous attempts to compare device performance of cylindrically radial and planar solar cells have focused on minority charge carrier diffusion in the quasi-neutral regions. As such, we do not attempt to improve upon previous efforts solving for minority charge carrier diffusion current densities in the quasi-neutral regions and, instead, focus on developing a generalized curriculum for the constituent components that govern solar cell *I-V* characteristics of non-planar geometries.

In the low injection limit, a *p-n* junction is typically considered to be spatially separable into two quasi-neutral regions and a space-charge region (SCR). We display a spatially generalized *p-n* junction in Fig. 1, with the distinct regions indicated by vector positions $\vec{r}_i, i = 0,1,2,3$, for the purpose of indicating the spatial orientation of the junction in this discussion. In addition, we also show proposed energy band modifications in the SCR, which we explain in further detail in subsequent sections. In our analysis, a given transport variable $X(\vec{r})$ in the *n*-type quasi-neutral region is implied by the relationship $X(\vec{r}_0 \leq \vec{r} \leq \vec{r}_1) \equiv X_N(\vec{r})$, based on the *p-n* junction shown in Fig. 1. Likewise, in the SCR, $X(\vec{r}_1 \leq \vec{r} \leq \vec{r}_2) \equiv X_{SC}(\vec{r})$, and in the *p*-type quasi-neutral region (PTR), $X(\vec{r}_2 \leq \vec{r} \leq \vec{r}_3) \equiv X_P(\vec{r})$.

## II. THEORY

### A. TOTAL DEVICE CURRENT

The area over which charge extraction occurs for a *p-n* junction is a function of position, and, therefore, is not necessarily uniform for non-planar devices. Therefore, current *density* (*i.e.* charge per unit time per unit area) is not necessarily conserved for all solar cell architectures, though certainly for the planar geometry [15-17, 20-24]. However, current (*i.e.* charge per unit time) is fundamentally conserved for all geometries. Therefore, we deviate from traditional methods attempting to model total current density of a solar cell and, instead, focus on calculating total current, because it is more fundamental to non-planar solar cell performance. To arrive at an expression for total current of the geometrically generalized solar cell device, we apply conservation of current at a specific position along the *p-n* junction; analogous to the methodology applied to planar solar cells. However, because manipulation of the drift-diffusion and continuity equations in the quasi-neutral regions yield expressions for current density, not current, we write conservation of current at a specific position in the *p-n* junction in terms of area integrals over current density; *i.e.*

$$i_{total} = \iint \vec{J}_n(\vec{r}) \cdot d\vec{a}(\vec{r}) + \iint \vec{J}_p(\vec{r}) \cdot d\vec{a}(\vec{r}). \qquad 1$$

According to the planar analysis [15-17, 20-24], appropriate positions along the solar cell device to sum the electron and hole current densities are at either $\vec{r} = \vec{r}_1$ or $\vec{r} = \vec{r}_2$, as these positions share boundaries with the SCR. By combining conservation of current density with a charge carrier continuity equation across the SCR, the expression for total current density becomes a sum of minority charge carrier current densities from the quasi-neutral regions, evaluated at the edges of the SCR, and generation and recombination current densities from across the SCR [15-17, 20-24]. Here, we employ the same methodology, but now apply it to conservation of current in Eq. 1. In addition, we re-write Eq. 1 in terms of generalized coordinates, so that the expression for total current may be utilized by any coordinate system. In this way, the expression for total current is universal for all geometrical orientations of a *p-n* junction, provided that the junction is established symmetrically along only one axis of a coordinate system (*i.e.* current density is flowing parallel to only one unit vector normal of an area element), and that the low-injection limit is applicable. For a three dimensional system of generalized coordinates $q_i$, the position vector $\vec{r}$ is defined by

$$\vec{r} = \sum_{i=1}^{3} q_i \hat{e}_i, \qquad 2$$

the gradient $\vec{\nabla}$ is defined by

$$\vec{\nabla} A = \sum_{i=1}^{3} \frac{1}{h_i(\vec{r})} \frac{\partial A}{\partial q_i} \hat{e}_i \qquad 3$$

for a scalar $A$, divergence is defined by

$$\vec{\nabla} \cdot \vec{B} = \frac{1}{h_1(\vec{r}) h_2(\vec{r}) h_3(\vec{r})} \left[ \frac{\partial}{\partial q_1} [h_2(\vec{r}) h_3(\vec{r}) B_1] + \frac{\partial}{\partial q_2} [h_1(\vec{r}) h_3(\vec{r}) B_2] \right. \\ \left. + \frac{\partial}{\partial q_3} [h_1(\vec{r}) h_2(\vec{r}) B_3] \right] \qquad 4$$

for a vector $\vec{B}$, and the infinitesimal area element $d\vec{a}(\vec{r})$ is defined by

$$da(\vec{r}) \hat{n} = \sum_{i=1}^{3} da_i(\vec{r}) \hat{e}_i. \qquad 5$$

For all expressions, the elements $\hat{e}_i$ represent unit vectors, and $h_i(\vec{r})$ represent the coordinate transformation factors (*e.g.* in cylindrical coordinates, $h_1(\vec{r}) = 1$, $h_2(\vec{r}) = \rho$, and $h_3(\vec{r}) = 1$).

For conservation of current evaluated at $\vec{r} = \vec{r}_1$, Eq. 1 becomes

$$i_{total} = \iint \vec{j}_{p_N}(\vec{r})\big|_{\vec{r}=\vec{r}_1} \cdot d\vec{a}(\vec{r})\big|_{\vec{r}=\vec{r}_1} + \iint \vec{j}_{n_N}(\vec{r})\big|_{\vec{r}=\vec{r}_1} \cdot d\vec{a}(\vec{r})\big|_{\vec{r}=\vec{r}_1}. \qquad 6$$

To determine the electron current density at $\vec{r} = \vec{r}_1$, ($\vec{j}_{n_N}(\vec{r})\big|_{\vec{r}=\vec{r}_1}$) in terms of the electron current density at $\vec{r} = \vec{r}_2$, ($\vec{j}_{n_P}(\vec{r})\big|_{\vec{r}=\vec{r}_2}$) we integrate the *electron* continuity equation across the SCR, assuming electron flow along only one coordinate axis $q_1$; *i.e.*

$$h_2(\vec{r}) h_3(\vec{r}) \vec{j}_{n_P}(\vec{r})\big|_{\vec{r}=\vec{r}_2} - h_2(\vec{r}) h_3(\vec{r}) \vec{j}_{n_N}(\vec{r})\big|_{\vec{r}=\vec{r}_1} \\ = -q \int_{\vec{r}_1}^{\vec{r}_2} h_1(\vec{r}) h_2(\vec{r}) h_3(\vec{r}) [G_{SC}(\vec{r}) - U_{SC}(\vec{r})] \, dq_1 \qquad 7$$

Combining Eq.'s 6 and 7, total current can be shown to be expressed as,

$$i_{total} = \iint \vec{J}_{p_N}(\vec{r})\big|_{\vec{r}=\vec{r}_1} \cdot d\vec{a}(\vec{r})|_{\vec{r}=\vec{r}_1} + \iint \frac{h_2(\vec{r})h_3(\vec{r})\vec{J}_{n_P}(\vec{r})\big|_{\vec{r}=\vec{r}_2}}{h_2(\vec{r})h_3(\vec{r})|_{\vec{r}=\vec{r}_1}} \cdot d\vec{a}(\vec{r})|_{\vec{r}=\vec{r}_1}$$
$$+ q \iiint_{r_1}^{r_2} [G_{SC}(\vec{r}) - U_{SC}(\vec{r})] \, d^3r. \quad\quad 8$$

Analogously, for current evaluated at $\vec{r} = \vec{r}_2$, total current becomes

$$i_{total} = \iint \frac{h_2(\vec{r})h_3(\vec{r})\vec{J}_{p_N}(\vec{r})\big|_{\vec{r}=\vec{r}_1}}{h_2(\vec{r})h_3(\vec{r})|_{\vec{r}=\vec{r}_2}} \cdot d\vec{a}(\vec{r})|_{\vec{r}=\vec{r}_2} + \iint \vec{J}_{n_P}(\vec{r})\big|_{\vec{r}=\vec{r}_2} \cdot d\vec{a}(\vec{r})|_{\vec{r}=\vec{r}_2}$$
$$+ q \iiint_{r_1}^{r_2} [G_{SC}(\vec{r}) - U_{SC}(\vec{r})] \, d^3r. \quad\quad 9$$

It is extremely important to note that Eq. 8 and Eq. 9 are physically identical, despite how they appear analytically. Upon being applied to a particular geometry (planar, cylindrical, and, spherical, as well as other less symmetric situations), either position will yield identical forms for the total current expression. Hence, these expressions are the most general form of total current for a *p-n* junction solar cell in the low injection limit. The equations for total current are expressed in Cartesian, cylindrical, and spherical coordinate systems in Appendix 1.

In order to evaluate Eq.'s 8 or 9 (*i.e.* in order to simulate *I-V* characteristics using Eq.'s 8 and 9), the solutions for the minority charge-carrier diffusion current-densities in the quasi-neutral regions ($\vec{J}_{p_N}(\vec{r}) = -qD_{p_N}\vec{\nabla}p_N(\vec{r})$ and $\vec{J}_{n_P}(\vec{r}) = qD_{n_P}\vec{\nabla}n_P(\vec{r})$), as well as the generation $G_{SC}(\vec{r})$ and recombination $U_{SC}(\vec{r})$ rates in the SCR, need to be predetermined. As mentioned earlier, we do not attempt to improve upon calculations of minority charge-carrier diffusion in the quasi-neutral regions for non-planar geometries. The equations governing low injection, minority charge-carrier diffusion in the quasi-neutral regions are still governed by

$$D_{\nu_\Omega} \nabla^2 \nu_\Omega(\vec{r}) - \frac{\Delta \nu_\Omega(\vec{r})}{\tau_{\nu_\Omega}} = -G_\Omega(\vec{r}'), \quad\quad 10$$

with $\nu_\Omega = p_N, n_P$ representing both quasi-neutral regions [15-17, 20-24]. In addition, voltage dependence is still obtained by the *law of the junction* (*i.e.* $n_\Omega(\vec{r})p_\Omega(\vec{r}) = n_i^2 exp(\beta qV)$), applied as a boundary condition for both of the quasi-neutral/space-charge interfaces [15-17, 20-24]. Because the law of the junction is a scalar condition, it is not affected by the geometry of the *p-n* junction. To analytically calculate total current of a solar cell device, solution(s) to Eq. 10 must exist for a given geometrical orientation of a *p-n* junction, which, at present, has only been achieved, analytically, in Cartesian coordinates using a one-dimensional approximation for charge transport. Changes to the expressions for the generation and recombination rates in the SCR are the subject of interest for the next two subsections.

## B. RECOMBINATION RATES

Because recombination rates $U_\Omega(\vec{r})$ in the quasi-neutral regions (in the low injection limit) are proportional to the difference between biased and un-biased minority charge carrier concentrations $\Delta \nu_\Omega(\vec{r})/\tau_{\nu_\Omega}$ [21], recombination is only implicitly spatially dependent. Therefore, there is no change in the explicit functional forms for radiative, Shockley-Reade-Hall (SRH), or Auger recombination (AR) rates in the quasi-neutral regions in a geometrically generalized solar cell device, as compared to the planar solar cell device. However, some recombination rates in the SCR are explicitly spatially dependent, and will take on different functional forms in different cell geometries. Appendix 2 highlights the geometrically generalized recombination rates considered here in the SCR, for the types of recombination discussed above [18]. Total recombination in the SCR is the sum of all recombination rates.

Because radiative recombination occurs between the conduction and valence bands, it is independent of position and, therefore, of geometry (this assumes uniform energy bands across the SCR). However, in the SCR, because energy is a function of position, both SRH and Auger recombination become spatially dependent. Sah, Noyce, and Shockley originally established the functional form for the spatial dependency [18] of SRH recombination in the SCR for a planar solar cell configuration. Their analysis incorporated spatial dependency into the energy of trap states within the band gap, and into the intrinsic chemical potential across the SCR of a planar, single junction solar cell. Based on the quadratic energy band curvature within the SCR (as solved for from Poisson's equation), they assumed a first-order, linear approximation for the energy dependency of the intrinsic chemical potential in the planar solar cell configuration [18]. However, for non-planar solar cell configurations, the solution to Poisson's equation in the SCR changes. While solutions still maintain a quadratic term, solutions to Poisson's equation in cylindrical and spherical geometries yield a logarithmic and hyperbolic lowest-order term, respectively, not a linear dependency, as the case is for the planar junction. Here, keeping only lowest order terms to approximate the spatial dependency of the intrinsic chemical potential, as Sah *et al.* did, we have adapted the analysis for SRH and Auger recombination across the SCR to account for cylindrical and spherical geometries. Our approach for determining the spatial dependence of the intrinsic chemical potential across the SCR is the basis for constructing the spatial dependence of the energy band diagram shown in Fig. 1 (*i.e.* linear for planar, logarithmic for cylindrical, and hyperbolic for spherical). Equations governing the intrinsic chemical potential for planar (Sah, *et al.*), cylindrical, and spherical geometries are indicated in Table 1.

**Table 1**. Spatial dependences of the intrinsic chemical potential within the space-charge region. Here, $V_{B.I.}$ represents the built-in bias of the junction.

| Geometry | Intrinsic Chemical Potential $\mu_i(\vec{r})$ in the SCR | |
|---|---|---|
| Planar | $\dfrac{\mu_i(z) - \dfrac{\varepsilon_{FC} + \varepsilon_{FV}}{2}}{q[V_{B.I.} - V]} = \dfrac{1}{r_2 - r_1}\left[z - \dfrac{r_1 + r_2}{2}\right]$ | 11 |
| Cylindrical | $\dfrac{\mu_i(\rho) - \dfrac{\varepsilon_{FC} + \varepsilon_{FV}}{2}}{q[V_{B.I.} - V]} = \dfrac{1}{\ln\left(\dfrac{r_2}{r_1}\right)} \ln\left(\dfrac{\rho}{\sqrt{r_1 r_2}}\right)$ | 12 |
| Spherical | $\dfrac{\mu_i(r) - \dfrac{\varepsilon_{FC} + \varepsilon_{FV}}{2}}{q[V_{B.I.} - V]} = -\dfrac{1}{r}\dfrac{r_1 r_2}{r_2 - r_1} + \dfrac{r_1}{r_2 - r_1} + \dfrac{1}{2}$ | 13 |

## C. GENERATION RATE

A one-dimensional expression for the charge-carrier generation rate $G(z')$ is available in any number of references on solar cell physics [15-17, 20-24]. This expression typically takes a form similar to

$$G(z') = \int_{\Delta}^{\varepsilon_{max}} [1 - \mathcal{R}(\varepsilon_\gamma)] \alpha(\varepsilon_\gamma) \frac{I_{AMX}(\varepsilon_\gamma)}{\varepsilon_\gamma} exp(-\alpha(\varepsilon_\gamma) z') d\varepsilon_\gamma, \qquad 14$$

where: $\varepsilon_\gamma$ represents photon energy, $\Delta$ is the energy band gap of the semiconductor, $\mathcal{R}(\varepsilon_\gamma)$ is the reflectance, $\alpha(\varepsilon_\gamma)$ is the absorption coefficient of the semiconductor, $z'$ is the path length for photon absorption in the semiconductor ($z' = r_3 - z$, for light entering through a p-type window layer), and $I_{AMX}(\varepsilon_\gamma)$ is the measured solar spectral irradiance, with $X = 0 \rightarrow 1.5$, for varying air mass values.

To generalize this charge generation rate, we retraced the constituent components in the one-dimensional derivation, taking into account all three spatial dimensions. The standard components of the generation rate derivation are 1) steady-state photon continuity, 2) Planck's law of radiation, 3) the Beer-Lambert law, 4) particle number conservation, and 5) the Earth-Sun orientation [15-24]. Combining these steps, we derive a geometrically generalized charge carrier generation rate as the divergence of a spatially decaying photon flux,

$$G(\vec{r'}) = -\vec{\nabla} \cdot \vec{\sigma}(\vec{r'}) \qquad 15$$

where the vector $\vec{\sigma}(\vec{r'})$ represents the spatially decaying incident photon flux (having units of $cm^{-2} s^{-1}$), defined by

$$\vec{\sigma}(\vec{r'}) = \hat{k} \int_{\Delta}^{\varepsilon_{max}} [1 - \mathcal{R}(\varepsilon_\gamma)] \frac{I_{AMX}(\varepsilon_\gamma)}{\varepsilon_\gamma} exp(-\vec{\alpha}(\varepsilon_\gamma) \cdot \vec{r'}) d\varepsilon_\gamma. \qquad 16$$

In this representation, the charge carrier generation rate takes the functional form of a typical source-term in fluid equations. The unit vector $\hat{k}$ defines the direction of incident light on the solar cell, and is defined by the vector representation of the speed of light $\vec{c} = c\hat{k}$. Equation 15 indicates that the generation rate of charge carriers is fundamentally dependent on incident light orientation with respect to the surface normal of the cell. This generalization appears to have yet been reported in the context of p-n junction solar cells. In the context of the photon continuity equation, $\vec{\sigma}(\vec{r'})$ is analogous to the Poynting vector in Poynting's Theorem [31]. In fact, for electromagnetic radiation being absorbed via the Beer-Lambert law, the only difference between $\vec{\sigma}(\vec{r'})$ and the Poynting vector $\vec{S}(\vec{r'})$, is that the Poynting vector describes the spatially decaying flux of the photon energy (with units of $eV \, cm^{-2} s^{-1}$) being absorbed, while $\vec{\sigma}(\vec{r'})$ describes the spatially decaying flux of the number of photons themselves. When applied to particular p-n junction orientations, Eq. 15 has unexpected consequences on the spatial dependences that the generation rate takes, which we discuss in detail in the Results and Discussion section.

### III. RESULTS AND DISCUSSION

**D. Space-Charge Region Recombination**

We have plotted the intrinsic chemical potential versus SCR position for planar, cylindrical, and spherical geometries in Fig. 2. From the plots in Fig. 2, it is seen that in addition to geometry of the junction, both SCR width ($r_2 - r_1$) and inner radius $r_1$ are spatial parameters that affect the curvature of the intrinsic chemical potential across the SCR. For the cylindrical and spherical architectures (Fig.'s 2b and 2c), it is seen that increasing the inner radius decreases the curvature of the intrinsic chemical potential, moving toward a linear spatial-dependence. This is indicative of a size-scale feature for non-planar architectures, where energy dependence across the SCR becomes more like that of a planar device. The dependence of the intrinsic chemical potential on the SCR width is opposite of that for the inner radius $r_1$, in that for increasingly larger space-charge regions, the energy dependence curvature increases. Here it is seen that for smaller space-charge regions, the intrinsic chemical potential is, again, approximately linear. The combination of both effects (increasing/decreasing inner radii and SCR widths) described above indicate that for SCR thicknesses that are small with respect to inner radius dimensions, recombination dynamics in the SCR can be approximated by planar device physics, as detailed by Sah, Noyce, and Shockley. This obvious conclusion about how geometry affects recombination in the SCR is mentioned here, only for verification of our results; *i.e.* it is a well-known fact that for many (if not all) systems, that the local physics on a thin cylindrical or spherical shell with a large radius can be approximated as planar. In simulating the recombination current across the SCR, however, appropriate geometries will still be needed to calculate the volume integrals (shown in Appendix 1). Previous models of non-planar, cylindrically radial *p-n* junctions used planar SRH recombination statistics in the SCR, as established by Sah, *et al.* [18]; valid only for a thin cylindrical shell with a large radius. Here, we are now able to model recombination statistics for planar, cylindrical, and spherical *p-n* junctions, for any ratio of SCR width to inner radius.

Inserting the spatial dependencies of the intrinsic chemical potentials from Table 1 into the recombination rates in Appendix 2, Fig. 3 shows the recombination statistics in the SCR as the sum of radiative, SRH, and Auger recombination. Note: SRH recombination dominates in the SCR by many orders of magnitude in the low injection limit. For SRH recombination, we assume that the energy of trap states in the band gap is equal to the intrinsic chemical potential. Fig. 3a reaffirms our choice for drawing the energy bands across the SCR in Fig. 1. From Fig. 3b, the non-linear intrinsic chemical potentials within the SCR have the effect of changing the position of the maximum recombination rate. Note, in Fig.'s 3a and 3b, the SCR distance is expressed logarithmically. In addition, our results for the planar recombination rate within the SCR closely match those reported by Sah, et *al* [18]. The maximum rate of recombination, for all geometries, occurs where the corresponding intrinsic chemical potential crosses the ordinate origin. Because of the spatial dependences of the intrinsic chemical potentials, the positions of maximum recombination within the SCR occur closer to $r_1$ for the cylindrical and spherical geometries.

Using Appendix 1 to calculate a recombination current across the SCR, we fix the volume for a planar slab, a cylindrical shell, and a hemispherical shell, and integrate the corresponding recombination rates over the volumes. Fig.'s 3c and 3d show dark current for the three geometries considered, with "short" carrier lifetimes ($10^{-8}$ second) and "long" carrier lifetimes ($10^{-6}$ second) in the SCR. Despite the fact that volume is constant, the recombination current produced from the SCR actually changes depending on which geometrical orientation the *p-n* junction takes. The results indicate there is an inextricable relationship connecting material properties and junction geometry concerning the performance of a solar cell. For the "short" carrier lifetimes, the spread in rectifying voltages for the three geometries is significantly larger than for the "long" carrier lifetimes. This trend indicates that for materials with "good" transport properties (*e.g.* lifetime, mobility, diffusivity, *etc.*), geometry affects overall solar cell performance less. We base this conclusion on the range of current shown and on typical short-circuit current values for a 1 $cm^2$ solar cell [15-24]. Based on these results, for *p-n* junctions that have significant contribution to total current from the SCR, the planar intrinsic chemical potential approximation across the SCR will yield increasingly invalid overestimates of dark current from the SCR.

E. **Generation Current**

To emphasize differences in the generation rates, as functions of both geometry and incident light orientation, short-circuit currents $I_{sc}$ are calculated for fixed-volume absorbers in planar, cylindrical, and hemispherical geometrical structures, by virtue of the integral equation

$$I_{sc}(\vec{r})|_{\vec{r}=\vec{R}} = q\, N(\vec{r})|_{\vec{r}=\vec{R}} \iiint G(\vec{r'}) \exp\left(-\frac{\vec{r}\cdot\hat{n}}{L_{diff}}\right) d^3r, \qquad 17$$

where $\exp\left(-\frac{\vec{r}\cdot\hat{n}}{L_{diff}}\right)$ is an ad hoc collection probability factor, $\hat{n}$ is the unit vector direction normal to the area element through which current is flowing, $N(\vec{r})|_{\vec{r}=\vec{R}}$ is the number of cells in a given area for the cylindrical and spherical architectures, and $L_{diff}$ is the average distance charge carriers will travel before a recombination event (*i.e.* the diffusion length). The notation $\vec{r}=\vec{R}$ is meant to denote that while the volume integral leaves no spatial variable dependence $\vec{r}$ in the short-circuit current, it is still dependent on spatial parameters $\vec{R}$ (*e.g.* radius, length) inherent to the geometry of the volume. The functional forms of the generation rates used for the various geometries and incident light orientations are shown in Appendix 3. In the architectural comparisons shown in Fig. 4, incident photon number onto all structures is conserved for all geometries, and 100% light absorption is assumed (this affects the number $N(\vec{r})|_{\vec{r}=\vec{R}}$ of cells per unit area). In addition, material properties (*e.g.* band gap, absorption coefficient, diffusion length, area, *etc.*) are held fixed across all architectures. It should be noted that for these results, the short-circuit current *density* that is calculated is taken from the view of the light source. By conserving the incident photon number onto each absorbing structure, we are able to obtain an expression for short-circuit current density, where the area element used to calculate it is the area of the cell that is exposed to light $A_{PV} = 1\, cm^2$; *i.e.* $j_{total} = {i_{total}}/{A_{PV}}$. For the cylindrical and hemispherical geometries, charge flow is taken to be entirely radial (*i.e.* $\hat{n} = \hat{\rho}$ and $\hat{n} = \hat{r}$, respectively).

Incident light does not naturally converge/concentrate in a radial fashion, though it may be engineered to do so with, for example, a microlens array [32]. Therefore, it is interesting to observe the functional differences in the generation rates produced from such effects. Fig. 4 shows short-circuit currents calculated via Eq. 17. For the cylindrical absorber with longitudinally incident light (*i.e.* $\hat{k} = -\hat{z}$), the short-circuit current shown is for a cylindrical absorber with a radius of 10 nm. Short-circuit currents for (a) 10 μm and (b) 100 nm diffusion lengths are shown together in Fig. 4 for comparison. While the peak of short-circuit current for the cylindrical absorber is only slightly greater than that of the planar absorber with a 10 μm diffusion length, the maximum cylindrical short-circuit current is almost two times the planar maximum for that of a 100 nm diffusion length. It is important to point out that the properties of current collection, based on values for diffusion length, are relative to the absorption properties of the material; for absorber thicknesses of approximately 500 nm, this hypothetical material has absorbed nearly 100% of all incident light, which can be seen from the cylindrical short-circuit current curves in Fig. 4. In general, for charge collection length scales that are small compared to average photon absorption depths, the cylindrical geometry has the potential to significantly increase short-circuit current, provided that the cylinder radius is small compared to recombination length scales of the material.

Also worthy of discussion is the fact that the cylindrical short-circuit current saturates to its maximum value for larger absorbing volumes. This is due to the fact that for longitudinally incident light with 100% light absorption, the cylindrical absorber volume only increases with cylinder length, not radius. Note, this will not necessarily be true when considering non-perfect absorption. Since the probability of current collection decreases only with increasing radius, nearly all photogenerated charge carriers will be extracted radially for the cylindrical absorber, despite any increase in cylinder length, provided that the cylinder radius is small compared to the charge carrier diffusion length (*i.e.* $r = 10\ nm \ll L_{diff} = 100\ nm$). Increases in cylinder length will only benefit the performance of short-circuit current by allowing for the absorption of more light. Therefore, as cylinder volume increases (*i.e.* as cylinder length increases), the short-circuit current plateaus because photon absorption has saturated, with nearly 100% of all incident photons having already been absorbed at a depth of approximately 500 nm. When directionally-dependent charge carrier collection $exp\left(-\frac{\vec{r}\cdot\hat{n}}{L_{diff}}\right)$ is considered in calculating the short-circuit current, the cylindrical absorber, with longitudinally incident light, is the only architecture to have this feature in short-circuit current, because it is the only situation where photon absorption and charge collection are mutually orthogonal ($\hat{\rho}\cdot\hat{z} = 0$).

However, it should also be noted that any increase in short-circuit current for the cylindrical absorber comes at a cost of more material. In Fig. 4b, the short-circuit current maximum for the cylindrical absorber occurs at almost an order of magnitude larger material volume, as compared to the maximum short-circuit current for the planar absorber. With this in mind, cost benefits will be necessary when considering non-planar architectures for "low" diffusion length materials.

Based on the short-circuit current calculated via Eq. 17, it is also possible to calculate the open-circuit voltage $V_{oc}$ and fill factor $FF$ of the absorber (see Appendix 4). From the short-circuit current, open-circuit voltage, and fill factor, the conversion output power $P$ is also determined by $P(\vec{r})|_{\vec{r}=\vec{R}} = I_{sc}(\vec{r})\ V_{oc}(\vec{r})\ FF(\vec{r})|_{\vec{r}=\vec{R}}$. These parameters are shown for the cylindrical absorber with longitudinally incident light in Fig. 5. The diffusion length used for Fig. 5 is 100 nm. In Fig. 5a, the open-circuit voltage is shown to decrease with both increasing cylinder radius and length. This decrease is more sensitive to radius, however, which can be accounted for from the fact that charge carrier collection probability only affects the radial direction. The decrease in open-circuit voltage with increasing length is accounted for from the fact that the saturation current $I_o$ of the absorber increases with cylinder length, which, itself, is a consequence of the exponentially decaying photon flux. The fill factor (Fig. 5b) also decreases with increasing cylinder radius and length; however, this effect is not as severe as compared to the open-circuit voltage. Fig. 5c shows the cylindrical behavior described in the previous paragraph, where the short-circuit current saturates at some maximum value for increasing cylinder length, but begins to decrease rapidly with increasing radius, as the collection probability becomes increasingly small for larger radii. The conversion output power is shown in Fig. 5d. Here, the output power decreases with increasing radius, accounting for smaller charge carrier collection probabilities with increasing radius. Unlike short-circuit current, however, output power reaches a local maximum with respect to length. This peak, with respect to length, arises from competing roles of absorption and saturation current (accounted for in the open-circuit voltage) with increasing cylinder length. The results show that the cylindrical absorber should be as thin as possible in the radial direction, and only long enough to sufficiently absorb light, while not diminishing performance with increasing dark (saturation) current. We do note that this conclusion, although derived from our calculations, has also been concluded experimentally [4] and through other radial *p-n* junction device simulations [26, 27, 30]. Results from Fig. 5 are similar to prior results reported for radial *p-n* junctions [25-30] that analyzed efficiency parameters according to cylinder length, based on solutions for minority charge carrier diffusion in the quasi-neutral regions. We have included the results in Fig. 5 to expand the spatial parameter space for cylindrically radial *p-n* junctions, by accounting for effects on efficiency parameters due to both cylinder radius and length, using a simpler model.

The open-circuit voltage, fill factor, and output power conversion are compared for planar, cylindrical, and hemispherical geometries, along with varying light incidences, in Figures 6 and 7, for diffusion lengths of 10 μm and 100 nm. Again, for longitudinally incident light onto the cylindrical absorber, a radius of 10 nm is used for these comparisons. The cylindrical absorber defines the upper limit, in terms of efficiency parameters ($I_{sc}, V_{oc}$, and $FF$), that all geometries and incident light orientations considered are capable of achieving. For a 100 nm diffusion length, the efficiency parameters are relatively the same at low material volumes. As material volume increases, the efficiency parameters for the cylindrical absorber maintain higher values than all other geometries and incident light orientations. As diffusion length increases, the efficiency parameters remain closely matched together for increasing material volumes.

For the material parameters considered in this simulation, at a 10 μm diffusion length, the peak planar output power is approximately the same as the peak cylindrical output power (Fig. 7a). Therefore, there appears to be little to no benefit of constructing solar cells in a non-planar architecture when the ratio of charge carrier collection length to average photon absorption depth is much greater than unity. However, in Fig. 7b, the cylindrical absorber clearly shows a performance enhancement over all other geometrical architectures. For 100% light absorption, the increase in material volume for the planar and cylindrical absorbers works out to be identical. However, for a 10 nm radius, the cylindrical absorber is capable of collecting charge much more efficiently than the planar device when light is incident longitudinally, and therefore, power conversion for the cylindrical architecture is significantly greater. Our results for device performance based on short-circuit current collected are precisely in sync with our results from calculations concerning recombination current from the SCR (*i.e.* non-planar geometries yielded less dark current in the SCR for semiconducting materials with "poor" transport properties, and showed little improvement over the planar geometry for semiconducting materials with "good" charge transport properties in the SCR). These results are in fairly good agreement with measured increases in device performance for amorphous silicon nanocoaxial solar cells [4]. While marked increases in short-circuit current have been reported for the nanocoaxial solar cell architecture, the open-circuit voltages have been shown not to be particularly affected. For comparable absorbing volumes (in the range of $10^{-6} - 10^{-5}$ $cm^3$), our results indicate that a cylindrical absorber shows improved efficiency primarily from increases in photon absorption, and thus by an increased short-circuit current, with little change in the open-circuit voltage. Therefore, based on a simple, yet effective, model employing a semi-empirical equation for short-circuit current, which has reciprocal relations to other solar cell performance parameters ($V_{oc}$ and $FF$), we conclude that when the ratio of charge collection length to average photon absorption depth is on the order of, or less than, unity, the cylindrical solar cell geometry is preferable over the planar and hemispherical solar cell geometries, to maximize efficiency.

While our results for device performance all stem from a semi-empirical expression for short-circuit current (Eq. 17), that has been adapted, here, to account for non-linear (*i.e.* radial) charge carrier transport in a solar cell, this simplistic model has been employed for the purpose of ascertaining important features of device performance for non-planar geometries, without having to solve the, cumbersome, second-order, non-linear, inhomogeneous, partial differential equations describing minority charge carrier diffusion, for which unrealistic approximations often have to be made, in order to solve them analytically (true, even more so, for the hemispherical architecture than the cylindrical architecture of a *p-n* junction solar cell). Because of this, it is not surprising that some of the features we have reported herein are similar in scope to those reported for cylindrically radial *p-n* junctions, as modeled by minority charge carrier diffusion in the quasi-neutral regions [26, 27]. However, the generalization of current, generation, and recombination, discussed here, provide a framework for comparing solar cells of all geometries, not only those of a planar and cylindrical symmetry. Moreover, this generalization is necessary to compare device performance as a function of geometry even for the simplistic model used here. Therefore, we conclude that this geometrical generalization of solar cell physics is crucial for understanding device performance as a function of geometry, even at the most basic level of device modeling.

## IV. CONCLUSION

We have derived analytical expressions for charge carrier transport in a geometrically generalized single *p-n* junction solar cell device in the low injection limit. In the low injection limit, the equations for total current and generation rate are universal for any solar cell geometry. Expressions for space-charge recombination have been derived based on spatially-dependent forms of the intrinsic chemical potential across the SCR. The expressions for total device current, generation rate, and recombination rates within the SCR, and the previously developed differential equations describing minority charge carrier diffusion in the quasi-neutral regions, outline a framework for performing detailed analytical calculations of *I-V* curves for planar, cylindrical, and spherical geometrical orientations of a *p-n* junction solar cell. Based on a simple, yet effective, model for determining the upper limit on short-circuit current, we conclude that when the ratio of charge collection length to average photon absorption depth is on the order of, or less than, unity, the cylindrical absorber (with light incident longitudinally) is the optimal geometry to construct a solar cell.

*Acknowledgments*

The authors would like to thank Professor Tonio Buonassisi for useful conversations regarding the energy spatial dependence across the space-charge region and for general help with the manuscript.

[32] Landy NI, Padilla WJ. Guiding light with conformal transformations. *Optics Express* 2009; **17**, No. 17, 14872.

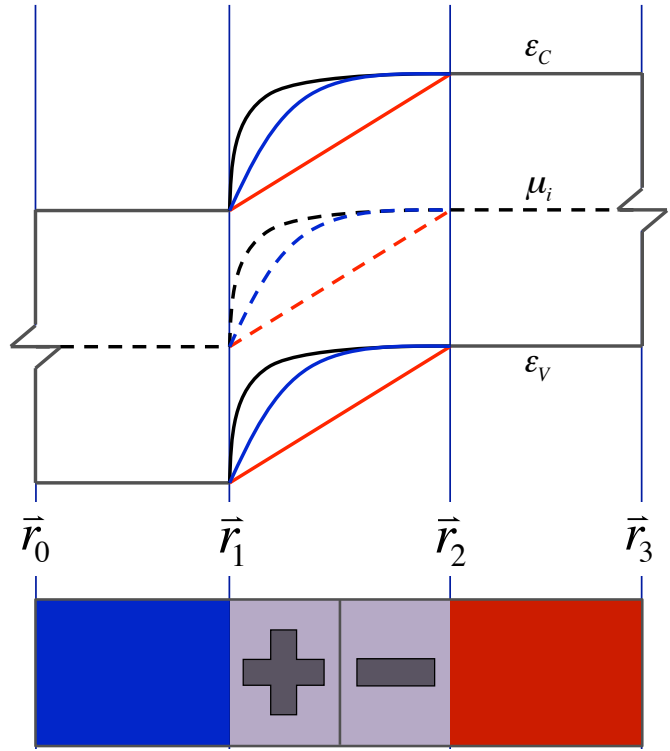

**FIG. 1**. Generalized *p-n* junction energy band diagram in the low injection limit. The spatial energy dependence is shown for planar (linear dependence; red), cylindrical (logarithmic dependence; blue), and spherical (inverse dependence; black) architectures.

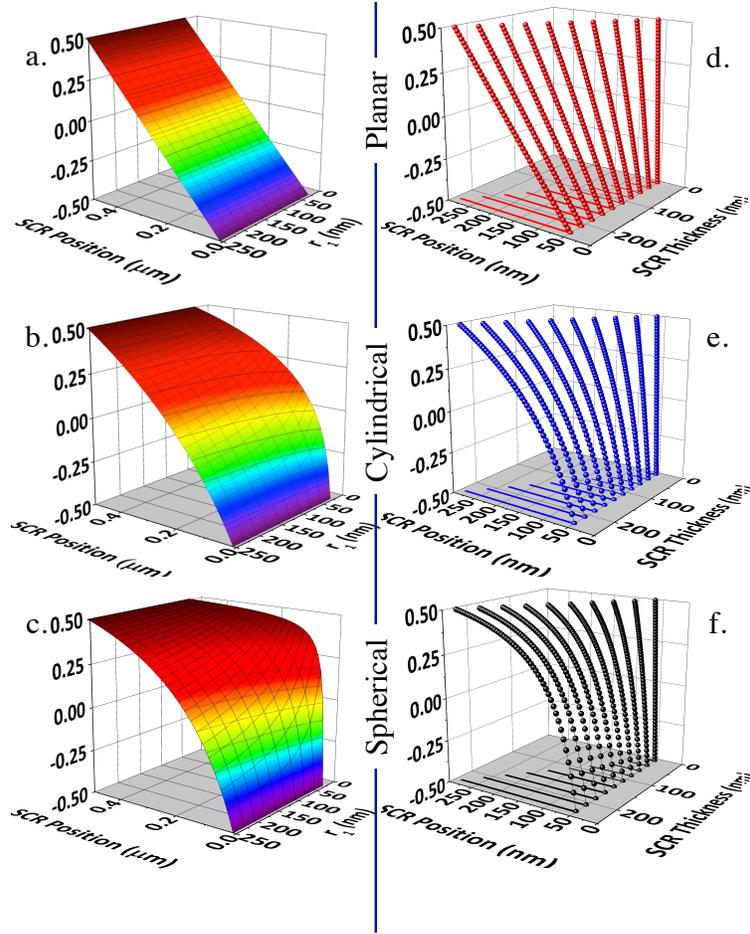

**FIG. 2**. Spatial dependence of the scaled intrinsic chemical potential $\frac{\mu_i(r)-b}{q[V_{B.I.}-V]}$ for the three solar cell architectures analyzed.

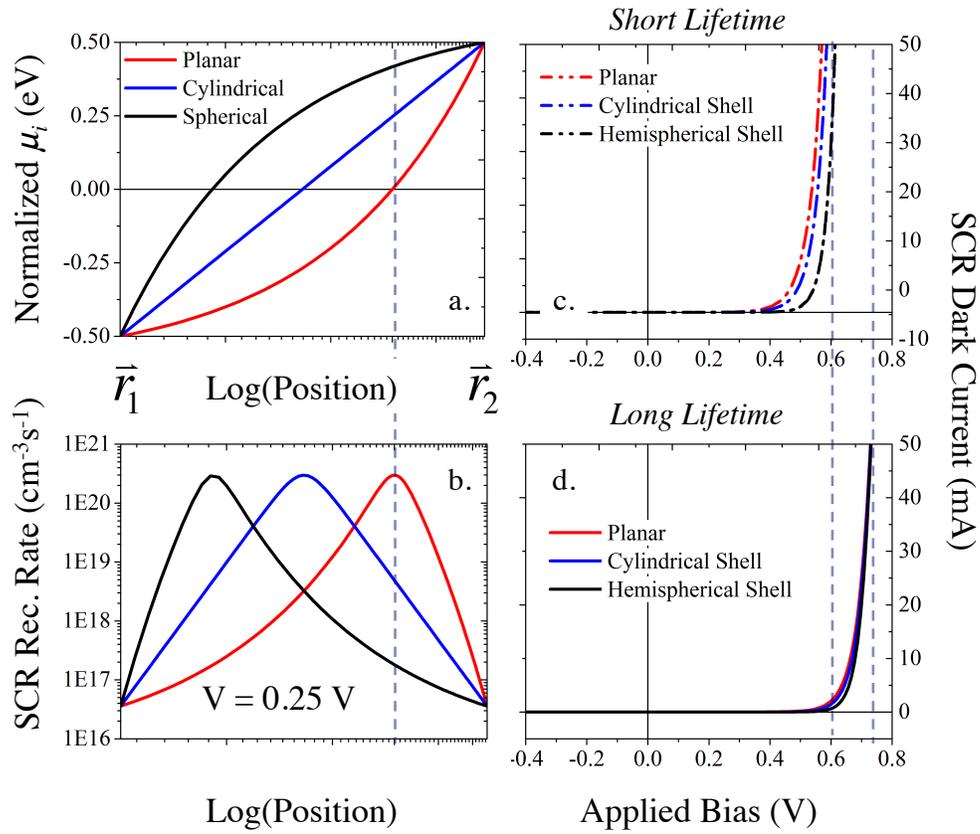

**FIG. 3**. (a) Intrinsic chemical potential as a function of logarithmic SCR position. (b) Recombination rates as a function of logarithmic SCR position. The overlaid dotted line indicates the midpoint of the SCR. Recombination (dark) current simulations for (c) $10^{-8}$ and (d) $10^{-6}$ charge carrier lifetimes in the SCR.

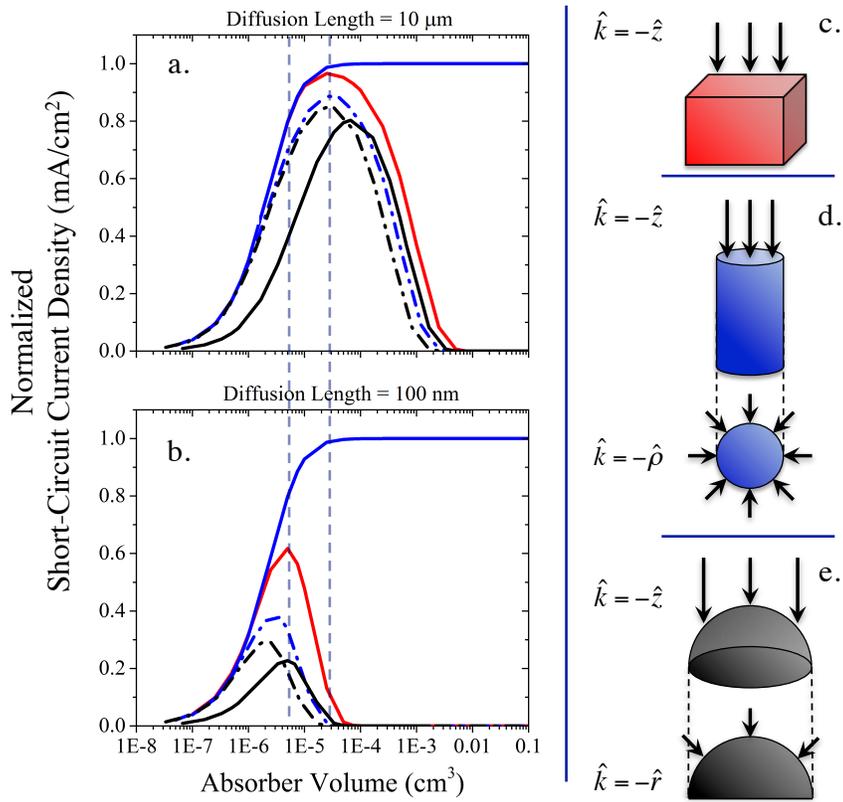

**FIG. 4**. Normalized short-circuit current density as a function of absorber volume for planar, cylindrical, and hemispherical architectures for (a) 10 µm and (b) 100 nm diffusion lengths. Longitudinally-incident and radially-incident light are indicated by solid and dashed lines, respectively. (c) Planar absorber. (d) Cylindrical absorber. (e) Hemispherical absorber.

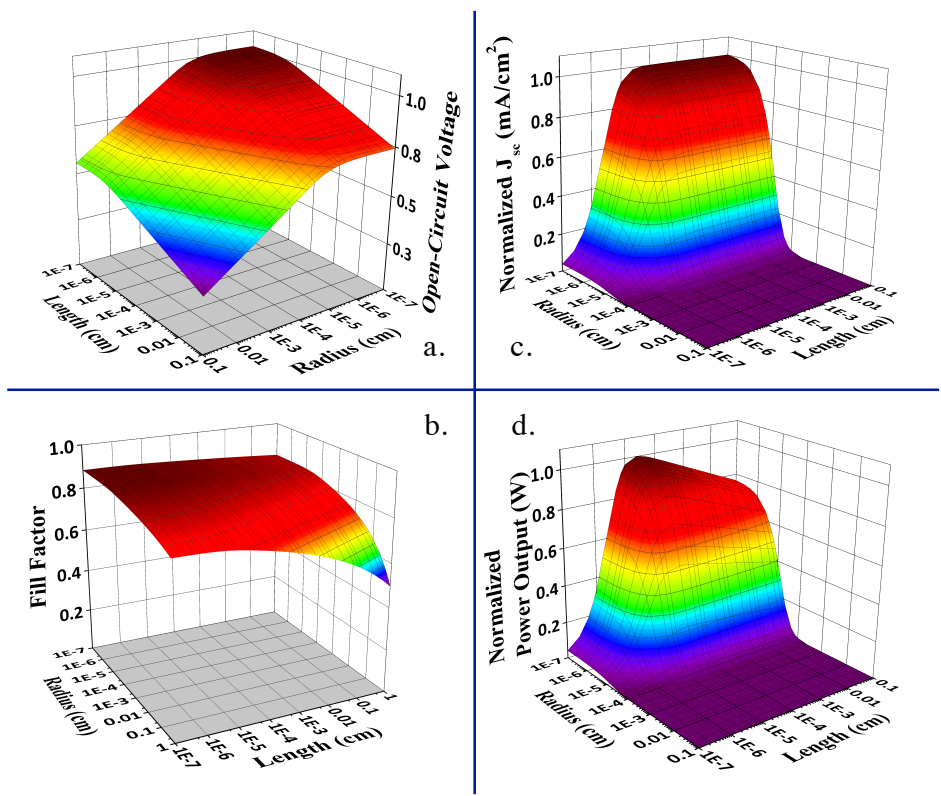

**FIG. 5**. (a) Open-circuit voltage, (b) fill factor, (c) normalized short-circuit current density, and (d) normalized output power for longitudinally-incident light on a cylindrical absorber as a function of cylindrical length and radius.

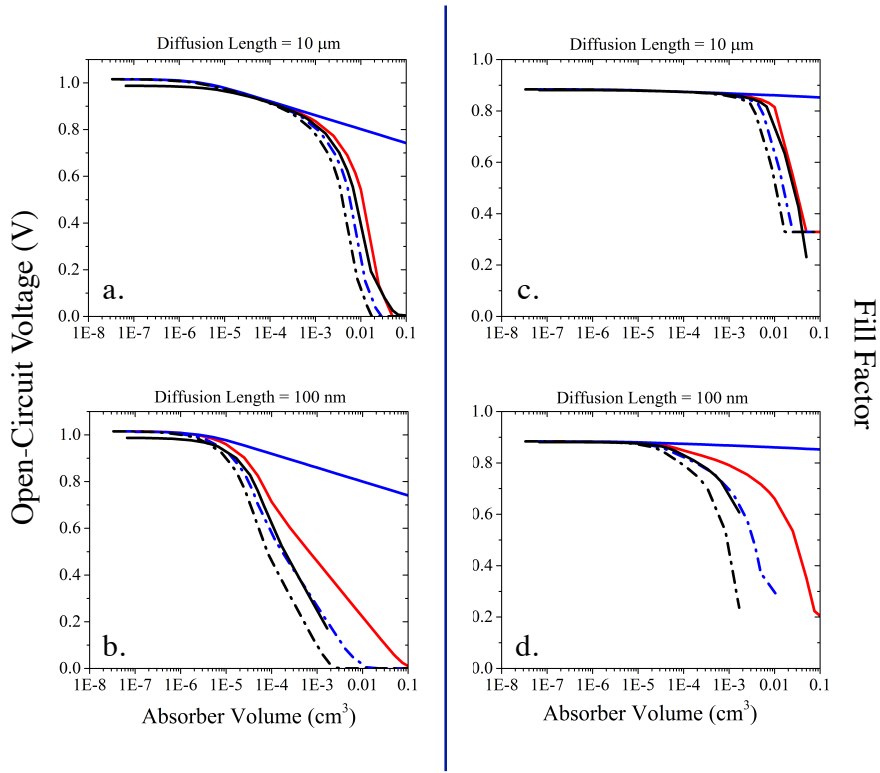

**FIG. 6**. Open-circuit voltages for (a) 10 μm and (b) 100 nm diffusion lengths, and fill factors for (c) 10 μm and (d) 100 nm diffusion lengths in Planar (red), cylindrical (blue), and hemispherical (black) geometries. Radially incident light orientations are indicated by dashed lines in these results.

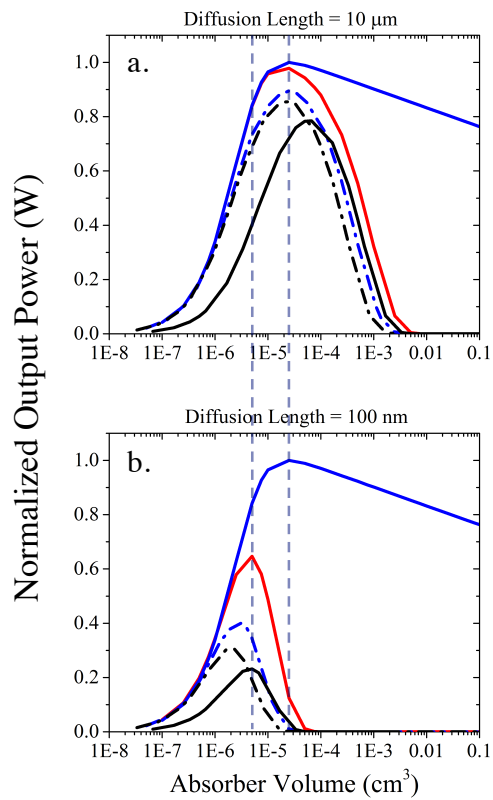

**FIG. 7**. Normalized output power for (a) 10 µm and (b) 100 nm diffusion lengths for planar (red), cylindrical (blue), and hemispherical (black) absorber architectures. Radially incident light is indicated by dashed lines in the plots.

**Appendix 1**.

Total current, calculated via Eq.'s 8 and 9, in Cartesian, and cylindrically and spherically symmetric geometries. In Cartesian coordinates, the expression for total current reduces to the conventional expression for planar solar cell current density, since the current densities, $j_{p_N}(z,x,y)\big|_{z=r_1}$ and $j_{n_P}(z,x,y)\big|_{z=r_2}$, are independent of the variables $x$ and $y$, and the area integrals over $x$ and $y$ are the (constant) cross sectional area $A_{PV}$ of the solar cell; $i.e.\ A_{PV} = \iint dxdy$. Prior calculations for total current *density* of cylindrically radial *p-n* junctions made approximations on where charge was being collected [27], since it is not a conserved quantity for non-planar *p-n* junction orientations.

| Geometrical Symmetry | Total Current Expression | |
|---|---|---|
| Planar | $i_{total} = \iint j_{p_N}(z,x,y)\big|_{z=r_1} dxdy + \iint j_{n_P}(z,x,y)\big|_{z=r_2} dxdy$ $+ q \iiint_{r_1}^{r_2} \left[G_{SC}(\vec{r'}) - U_{SC}(\vec{r})\right] dzdxdy$ | i |
| Cylindrical | $i_{total} = r_1 \iint j_{p_N}(\rho,\phi,z)\big|_{\rho=r_1} d\phi dz + r_2 \iint j_{n_P}(\rho,\phi,z)\big|_{\rho=r_2} d\phi dz$ $+ q \iiint_{r_1}^{r_2} \left[G_{SC}(\vec{r'}) - U_{SC}(\vec{r})\right] \rho d\rho d\phi dz$ | ii |
| Spherical | $i_{total} = r_1^2 \iint j_{p_N}(r,\theta,\phi)\big|_{r=r_1} \sin(\theta) d\theta d\phi$ $+ r_2^2 \iint j_{n_P}(r,\phi,z)\big|_{r=r_2} \sin(\theta) d\theta d\phi$ $+ q \iiint_{r_1}^{r_2} \left[G_{SC}(\vec{r'}) - U_{SC}(\vec{r})\right] r^2 \sin(\theta) dr d\theta d\phi$ | iii |

**Appendix 2**.

Geometrically generalized expressions for (iv) Radiative, (v) Shockley-Reade-Hall, and (vi) Auger recombination within the SCR of a homojunction device.

| Rate | Space-Charge Region (SCR) Recombination | |
|---|---|---|
| Radiative | $U_{Rad} = B\, n_i^2 [\exp(\beta qV) - 1]$ | iv |
| S.R.H. | $U_{SRH}(\vec{r}) = \dfrac{\dfrac{n_i}{\sqrt{\tau_{p_{SC}}\tau_{n_{SC}}}} \sinh\left(\dfrac{\beta qV}{2}\right)}{\exp\left(-\dfrac{\beta qV}{2}\right)\cosh\left(\beta[\varepsilon_t(\vec{r}) - \mu_i(\vec{r})] - \ln\left(\sqrt{\dfrac{\tau_{n_{SC}}}{\tau_{p_{SC}}}}\right)\right) + \cosh\left(\beta\left[\mu_i(\vec{r}) - \dfrac{\varepsilon_{FC} + \varepsilon_{FV}}{2}\right] + \ln\left(\sqrt{\dfrac{\tau_{p_{SC}}}{\tau_{n_{SC}}}}\right)\right)}$ | v |
| Auger | $U_{Aug}(\vec{r}) = 4n_i^3 \sqrt{\Lambda_{p_{SC}}\Lambda_{n_{SC}}} \cosh\left(\beta[\mu_i(\vec{r}) - b] + \ln\left(\sqrt{\dfrac{\Lambda_{p_{SC}}}{\Lambda_{n_{SC}}}}\right)\right) \exp(\beta qV) \sinh\left(\dfrac{\beta qV}{2}\right)$ | vi |

We derive expressions for SRH and Auger recombination rates analogous to the method performed by Sah, Noyce, and Shockley [18]. The explicit spatial dependence of these recombination rates is shown implicitly in terms of the intrinsic chemical potential $\mu_i(\vec{r})$. In our analysis of SRH recombination, we assume mid-gap trap states, which we approximate to be equivalent to the intrinsic chemical potential energy level across the SCR. In order to evaluate these recombination rates, it is necessary to have an explicit spatial dependence for the intrinsic chemical potential for a given geometrical orientation of the *p-n* junction.

**Appendix 3**.

Functional expressions for generation rate in planar (vii), cylindrical, and hemispherical geometries, and for longitudinal (viii and x) and radial light incidence (ix and xi).

| Geometry | $\hat{n}$ | $N(\vec{r})\|_{\vec{r}=\vec{R}}$ | Generation Rate; $-\vec{\nabla} \cdot \vec{S}(\vec{r'})$ | |
|---|---|---|---|---|
| Planar $\hat{k} = -\hat{z}$ | $\hat{z}$ | 1 | $\int_{\Delta}^{\varepsilon_{max}} [1 - \mathcal{R}(\varepsilon_\gamma)] \frac{I_{AMX}(\varepsilon_\gamma)}{\varepsilon_\gamma} \alpha(\varepsilon_\gamma) \exp\left(-\alpha(\varepsilon_\gamma)\left[\frac{Volume}{A_{PV}} - z\right]\right) d\varepsilon_\gamma$ | vii |
| Cylindrical $\hat{k} = -\hat{z}$ | $\hat{\rho}$ | $\frac{A_{PV}}{\pi R^2}$ | $\int_{\Delta}^{\varepsilon_{max}} [1 - \mathcal{R}(\varepsilon_\gamma)] \frac{I_{AMX}(\varepsilon_\gamma)}{\varepsilon_\gamma} \alpha(\varepsilon_\gamma) \exp(-\alpha(\varepsilon_\gamma)[L - z]) d\varepsilon_\gamma$ | viii |
| Cylindrical $\hat{k} = -\hat{\rho}$ | $\hat{\rho}$ | $\frac{A_{PV}}{2\pi R L}$ | $\int_{\Delta}^{\varepsilon_{max}} [1 - \mathcal{R}(\varepsilon_\gamma)] \frac{I_{AMX}(\varepsilon_\gamma)}{\varepsilon_\gamma} \alpha(\varepsilon_\gamma) \exp(-\alpha(\varepsilon_\gamma)[R - \rho]) d\varepsilon_\gamma$ | ix |
| Spherical $\hat{k} = -\hat{z}$ | $\hat{r}$ | $\frac{A_{PV}}{\pi R^2}$ | $\cos^2(\theta) \int_{\Delta}^{\varepsilon_{max}} [1 - \mathcal{R}(\varepsilon_\gamma)] \frac{I_{AMX}(\varepsilon_\gamma)}{\varepsilon_\gamma} \alpha(\varepsilon_\gamma) \exp\left(-\alpha(\varepsilon_\gamma)[R$ | x |
| Spherical $\hat{k} = -\hat{r}$ | $\hat{r}$ | $\frac{A_{PV}}{2\pi R^2}$ | $\int_{\Delta}^{\varepsilon_{max}} [1 - \mathcal{R}(\varepsilon_\gamma)] \frac{I_{AMX}(\varepsilon_\gamma)}{\varepsilon_\gamma} \alpha(\varepsilon_\gamma) \exp(-\alpha(\varepsilon_\gamma)[R - r]) d\varepsilon_\gamma$ | xi |

**Appendix 4**.

Reciprocal relations of open-circuit voltage and fill factor with short-circuit current. Open-circuit voltage is determined from short-circuit current via the following relation [15-17, 20-24],

$$V_{oc}(\vec{r})|_{\vec{r}=\vec{R}} = \frac{1}{q\beta} \ln\left(\frac{I_{sc}(\vec{r})}{I_o(\vec{r})} + 1\right)\bigg|_{\vec{r}=\vec{R}} \qquad \text{xii}$$

with $I_o$ representing the saturation current of the solar cell, defined, here, by

$$I_o(\vec{r})|_{\vec{r}=\vec{R}} = q\, B\, n_i^2\, V_{total}(\vec{r})|_{\vec{r}=\vec{R}}. \qquad \text{xiii}$$

In these expressions, $\beta = (k_B T)^{-1}$, $n_i$ represents the intrinsic charge carrier concentration, $V_{total}$ represents the total volume of the absorber, and $B$ is a constant of radiative recombination having units of $cm^3/s$. From Eq. xii, the fill factor can also be determined via the expression [15-17, 20-24],

$$FF(\vec{r})|_{\vec{r}=\vec{R}} = \frac{\tilde{V}_{oc}(\vec{r}) - \ln(\tilde{V}_{oc}(\vec{r}) + 0.72)}{\tilde{V}_{oc}(\vec{r}) + 1}\bigg|_{\vec{r}=\vec{R}} \qquad \text{xiv}$$

with $\tilde{V}_{oc}(\vec{r})$ given by [15-17, 20-24],

$$\tilde{V}_{oc}(\vec{r})\big|_{\vec{r}=\vec{R}} = q\, \beta\, V_{oc}(\vec{r})|_{\vec{r}=\vec{R}}. \qquad \text{xv}$$